\documentclass[preprint]{ptephy_om}

\preprintnumber{KYUSHU-HET-307, RIKEN-iTHEMS-Report-25} 

\usepackage{graphics}

\usepackage{amsmath} 
\usepackage{amsthm} 
\usepackage{url} 
\usepackage{stmaryrd}
\usepackage{tikz}
\usepackage{physics}
\usepackage{bm}
\usepackage{booktabs}
\usepackage{subcaption}

\numberwithin{equation}{section}

\allowdisplaybreaks

\begin{document}

\title{Monte Carlo Simulation of the $SU(2)/\mathbb{Z}_2$ Yang--Mills Theory}


\author[1]{Motokazu Abe}
\affil[1]{Department of Physics, Kyushu University, 744 Motooka, Nishi-ku,
Fukuoka 819-0395, Japan}

\author{Okuto Morikawa}
\affil{Interdisciplinary Theoretical and Mathematical Sciences Program
(iTHEMS), RIKEN, Wako 351-0198, Japan}

\author[1]{Hiroshi Suzuki}





\begin{abstract}%
We carry out a hybrid Monte Carlo (HMC) simulation of the $SU(2)/\mathbb{Z}_2$
Yang--Mills theory in which the $\mathbb{Z}_N$ 2-form flat gauge field (the
't~Hooft flux) is explicitly treated as one of the dynamical variables. We
observe that our HMC algorithm in the $SU(2)/\mathbb{Z}_2$ theory drastically
reduces autocorrelation lengths of the topological charge and of a physical
quantity which couples to slow modes in the conventional HMC simulation of the
$SU(2)$ theory. Provided that sufficiently large lattice volumes are available,
therefore, the HMC algorithm of the $SU(N)/\mathbb{Z}_N$ theory could be
employed as an alternative for the simulation of the $SU(N)$ Yang--Mills
theory, because local observables are expected to be insensitive to the
difference between $SU(N)$ and~$SU(N)/\mathbb{Z}_N$ in the large volume limit.
A possible method to incorporate quarks [fermions in the fundamental
representation of~$SU(N)$ with the baryon number~$1/N$] in this framework is
also considered.
\end{abstract}

\subjectindex{B01,B03,B31}

\maketitle

\section{Introduction}
\label{sec:1}
Recently, consideration on the Yang--Mills partition function with the
't~Hooft flux~\cite{tHooft:1979rtg} has
revived~\cite{Kitano:2017jng,Tanizaki:2022ngt,Nguyen:2023fun}, largely
motivated by the perspective of the generalized
symmetries~\cite{Gaiotto:2014kfa}, in particular in connection with the study
in~Ref.~\cite{Gaiotto:2017yup}. See also recent related
studies~\cite{Yamazaki:2017ulc,Hayashi:2023wwi,Hayashi:2024qkm,Hayashi:2024yjc,Hayashi:2024psa}, in which the 't~Hooft flux plays the crucial role in analyses
of the low-energy dynamics of gauge theory.

In the present paper, partially motivated by the above developments and
partially motivated by the possibility of a geometrical definition of the
fractional topological charge~\cite{Abe:2023ncy} in the $SU(N)$ Yang--Mills
theory with the 't~Hooft flux~\cite{tHooft:1981nnx,vanBaal:1982ag}, we carry
out a hybrid Monte Carlo (HMC) lattice simulation of the $SU(2)/\mathbb{Z}_2$
Yang--Mills theory, in which the 't~Hooft flux, the 2-form gauge field coupled
to the $\mathbb{Z}_2$ 1-form symmetry in the modern language, is dynamical.
Traditionally, lattice simulations of the $SU(N)/\mathbb{Z}_N$ Yang--Mills
theory are carried out by adopting the Wilson plaquette action in the adjoint
representation, which is blind on the
center~$\mathbb{Z}_N$~\cite{Halliday:1981te,Creutz:1982ga}; in this way,
nontrivial $SU(N)/\mathbb{Z}_N$ bundle structures are effectively summed over.
See also Refs.~\cite{Edwards:1998dj,deForcrand:2002vs}. In this paper, instead,
we treat the $\mathbb{Z}_N$ 2-form flat gauge field explicitly as one of the
dynamical variables; therefore, each gauge field configuration in the Monte
Carlo simulation explicitly possesses the value of the $\mathbb{Z}_N$ 2-form
gauge field. In this sense, our idea is similar in spirit to the study
in~Ref.~\cite{Kovacs:2000sy}, in which the $SU(2)$ Yang--Mills partition
function with a given fixed 't~Hooft flux is computed, although we make the
't~Hooft flux dynamical; see also Ref.~\cite{Halliday:1981tm} for an earlier
Monte Carlo study with an explicit dynamical $\mathbb{Z}_N$ 2-form gauge field.
Here, we adopt (a variant of) the HMC simulation algorithm~\cite{Duane:1987de}
having future applications of our method to lattice quantum chromodynamics
(QCD) [$SU(3)$ Yang--Mills theory with fundamental quarks] in mind.

Meanwhile, for many years, it has been known that the conventional HMC
algorithm on the periodic lattice suffers from the topological freezing
problem, a phenomenon whereby the Monte Carlo update is stuck within a
particular topological sector toward the continuum
limit~\cite{DelDebbio:2002xa,Schaefer:2010hu}; see Ref.~\cite{Bonanno:2024zyn}
and references therein for recent analyses. The HMC simulation with open
boundary conditions~\cite{Luscher:2011kk}, by activating in/out flows of
topological charges from lattice boundaries, appears to remove topological
barriers in between otherwise topological sectors and solve this problem. Here,
we see an analogue with the HMC simulation in the $SU(N)/\mathbb{Z}_N$ gauge
theory, in which the topological charge is shuffled by random dynamics of the
$\mathbb{Z}_N$ 2-form gauge field.

In continuum theory, the topological charge~$Q$ in the $SU(N)$ gauge theory
on a 4-torus, $T^4$, is defined by\footnote{Throughout this paper, we take a
convention that the field strength~$F_{\mu\nu}(x)$ is a hermitian matrix.}
\begin{equation}
   Q:=\int_{T^4}d^4x\,\frac{1}{32\pi^2}\varepsilon_{\mu\nu\rho\sigma}
   \tr\left[F_{\mu\nu}(x)F_{\rho\sigma}(x)\right].
\label{eq:(1.1)}
\end{equation}
In the $SU(N)$ Yang--Mills theory, $Q\in\mathbb{Z}$. However, in the
$SU(N)/\mathbb{Z}_N$ theory, $Q$ can be \emph{fractional\/}
as~\cite{tHooft:1981nnx,vanBaal:1982ag}
\begin{equation}
   Q=-\frac{1}{N}\frac{\varepsilon_{\mu\nu\rho\sigma}B_{\mu\nu}B_{\rho\sigma}}{8}
   +\mathbb{Z},
\label{eq:(1.2)}
\end{equation}
where $B_{\mu\nu}\in\mathbb{Z}$ is the 't~Hooft flux. Since $B_{\mu\nu}$ takes
discrete values, we make a random choice of $B_{\mu\nu}$ in the HMC simulation of
$SU(N)/\mathbb{Z}_N$ theory (see below). Then, this random choice inevitably
shuffles the topological charge~$Q$ and we may expect that the topological
sectors are efficiently sampled. We will observe that this expectation is
actually true.

This paper is organized as follows. In~Section~\ref{sec:2}, we briefly
summarize basic facts about the $SU(N)/\mathbb{Z}_N$ theory, mainly to set up
our lattice formulation on the periodic lattice of size~$L$,
$\Gamma:=(\mathbb{Z}/L\mathbb{Z})^4$. Under a certain gauge fixing of the
$\mathbb{Z}_N$ 1-form gauge symmetry, the 2-form $\mathbb{Z}_N$ gauge field
takes a particular form~\eqref{eq:(2.10)} of the ``$B$-field'' over which we
carry out the ``functional integral.'' Section~\ref{sec:3} is the main part of
this paper. Since the $B$-field in~Eq.~\eqref{eq:(2.10)} takes values
in~$\mathbb{Z}$ and has no obvious conjugate momentum, we have to appropriately
modify the HMC algorithm which is based on the molecular dynamics (MD) of
continuous variables. In~Section~\ref{sec:3.1}, we present our
``halfway-updating'' HMC algorithm which fulfills the detailed balance. The
proof of the detailed balance is deferred to~Appendix~\ref{sec:A}.
In~Section~\ref{sec:3.2}, we study the autocorrelation function of the
topological charge in the HMC history; we compare it in the $SU(2)$ theory
(i.e., without the $B$-field) with the one in the $SU(2)/\mathbb{Z}_2$
theory (with the $B$-field). For the lattice topological charge, we employ the
tree-level improved definition in~Ref.~\cite{Alexandrou:2015yba}, in which the
lattice field smeared by the gradient flow~\cite{Luscher:2010iy} is
substituted. We observe a drastic reduction of the autocorrelation in the HMC
simulation in the latter theory as anticipated above. We also observe a similar
reduction of the autocorrelation in the ``energy-operator'' $E(t)$ defined by
the gradient flow~\cite{Luscher:2010iy}. As shown in~Section~\ref{sec:2}, the
difference between $SU(N)/\mathbb{Z}_N$ and~$SU(N)$ can be understood as the
difference in boundary conditions (and the sum over them) and thus local
observables are expected to be insensitive to the difference between $SU(N)$
and~$SU(N)/\mathbb{Z}_N$ in the large volume limit in these gapped theories. To
have some idea on this point, we carry out an exploratory study on the finite
size effect in the continuum extrapolation of the topological susceptibility.
In~Section~\ref{sec:4}, we present a possible method to incorporate fermions in
the fundamental representation of~$SU(N)$ (quarks) in this framework. This is
achieved by gauging the baryon number~$U(1)$, $U(1)_B$, and embedding
$\mathbb{Z}_N$ into $SU(N)\times U(1)_B$. The $U(1)_B$ gauge boson unwanted for
QCD is made super-heavy by the St\"uckelberg mechanism on lattice.
Section~\ref{sec:5} is devoted to a conclusion. In~Appendix~\ref{sec:B}, we
present HMC histories and histograms of the topological charge for various
lattice parameters.

\section{$SU(N)/\mathbb{Z}_N$ Yang--Mills theory on~$T^4$}
\label{sec:2}
The $SU(N)/\mathbb{Z}_N$ Yang--Mills theory on~$T^4$ of size~$L$ in continuum
can be defined as follows. See~Refs.~\cite{Kapustin:2014gua,Tanizaki:2022ngt}.
We define boundary conditions of the $SU(N)$ gauge potential 1-form $a$
on~$T^4$ by
\begin{equation}
   a(x+L\Hat{\mu})
   =g_\mu(x)^\dagger a(x)g_\mu(x)-ig_\mu(x)^\dagger\mathrm{d}g_\mu(x),
\label{eq:(2.1)}
\end{equation}
where $x_\mu=0$ and $\Hat{\mu}$ is the unit vector in the $\mu$~direction; the
transition functions~$g_\mu(x)$ are $SU(N)$-valued. Here, since the gauge
potential behaves as the adjoint representation under the gauge transformation,
one may relax the cocycle condition for the transition functions
at~$x_\mu=x_\nu=0$ by $\mathbb{Z}_N$ factors as
\begin{equation}
   g_\mu(x)g_\nu(x+L\Hat{\mu})g_\mu(x+L\Hat{\nu})^\dagger g_\nu(x)^\dagger
   =e^{2\pi i B_{\mu\nu}(x)/N}\bm{1},
\label{eq:(2.2)}
\end{equation}
where $B_{\mu\nu}(x)\in\mathbb{Z}$ and $B_{\mu\nu}(x)=-B_{\nu\mu}(x)$. For the
consistency of transition functions among ``quadruple'' overlaps, it is
required that $B_{\mu\nu}(x)$ is flat modulo~$N$, $\mathrm{d}B=0\bmod N$. This
defines the $SU(N)/\mathbb{Z}_N$ principal bundle and the $SU(N)/\mathbb{Z}_N$
Yang--Mills theory over~$T^4$. It can be shown that we may take constant
$B_{\mu\nu}(x)$, $B_{\mu\nu}$. These integers are called the 't~Hooft fluxes.

It is well-understood how to implement this gauge theory on~$T^4$ as a lattice
gauge theory~\cite{Mack:1978kr,Ukawa:1979yv,Seiler:1982pw}; see also
Appendix~A.4 of~Ref.~\cite{Tanizaki:2022ngt} for a nice exposition. One first
introduces link variables as~$\Tilde{U}(x,\mu)\sim\exp(i\int_x^{x+\Hat{\mu}}a)$.
Boundary conditions on~$T^4$ are then defined by, as a lattice counterpart
of~Eq.~\eqref{eq:(2.1)},\footnote{Here is a side remark on the Wilson line
wrapping around a cycle of~$T^4$: Under the ordinary 0-form gauge
transformation,
\begin{equation}
   \Tilde{U}(x,\nu)\to
   \Omega(x)^\dagger\Tilde{U}(x,\nu)\Omega(x+\Hat{\nu}),
\label{eq:(2.3)}
\end{equation}
where $\Omega(x)$ is not necessarily periodic, the transition functions are
transformed as
\begin{equation}
   g_\mu(x)\to
   \Omega(x)g_\mu(x)\Omega(x+L\Hat{\mu})^\dagger
\label{eq:(2.4)}
\end{equation}
and one sees that the cocycle condition~\eqref{eq:(2.2)} is invariant under
this. The Wilson line of~$\Tilde{U}(x,\mu)$ thus should not contain the
transition functions at the boundary, because under~Eq.~\eqref{eq:(2.3)},
\begin{align}
   &\dotsb\Tilde{U}(x+(L-1)\Hat{\mu},\mu)\Tilde{U}(x+L\Hat{\mu},\mu)\dotsb
\notag\\
   &\to\dotsb\Omega(x+(L-1)\Hat{\mu})^\dagger
   \Tilde{U}(x+(L-1)\Hat{\mu},\mu)
   \Tilde{U}(x+L\Hat{\mu},\mu)
   \Omega(x+(L+1)\Hat{\mu})\dotsb.
\label{eq:(2.5)}
\end{align}
}
\begin{equation}
   \Tilde{U}(x+L\Hat{\mu},\nu)
   =g_\mu(x)^\dagger\Tilde{U}(x,\nu)g_\mu(x+\Hat{\nu}),
\label{eq:(2.6)}
\end{equation}
where $x_\mu=0$. We first regard link variables with~$x_\rho=L$ for a
certain~$\rho$ are all expressed by link variables with~$x_\rho=0$ by the
boundary conditions~\eqref{eq:(2.6)}. Then, we make the change of link
variables from $\Tilde{U}\to U$ by
\begin{equation}
   \Tilde{U}(x,\mu)=
   \begin{cases}
   U(x,\mu)g_\mu(x)&\text{for $x_\mu=L-1$},\\   
   U(x,\mu)&\text{otherwise}.\\
   \end{cases}
\label{eq:(2.7)}
\end{equation}
The new variables $U$ are regarded as obeying periodic boundary conditions,
$U(x+L\Hat{\mu},\nu)=U(x,\nu)$, where $x_\mu=0$. Under this change of variables,
one finds that the Boltzmann weight defined by the Wilson plaquette action
acquires $\mathbb{Z}_N$ factors as (letting $\beta$ the lattice bare gauge
coupling),
\begin{align}
   &\exp\left\{
   \beta\sum_{x\in\Gamma}\sum_{\mu<\nu}
   \frac{1}{N}\Re\tr\left[\Tilde{P}(x,\mu,\nu)-\bm{1}\right]
   \right\}
\notag\\
   &=
   \exp\left\{
   \beta\sum_{x\in\Gamma}\sum_{\mu<\nu}
   \frac{1}{N}\Re\tr\left[e^{-2\pi iB_{\mu\nu}(x)/N}P(x,\mu,\nu)-\bm{1}\right]
   \right\},
\label{eq:(2.8)}
\end{align}
where plaquette variables are
\begin{align}
   \Tilde{P}(x,\mu,\nu)&:=\Tilde{U}(x,\mu)\Tilde{U}(x+\Hat{\mu},\nu)
   \Tilde{U}(x+\Hat{\nu},\mu)^\dagger\Tilde{U}(x,\nu)^\dagger,
\notag\\
   P(x,\mu,\nu)&:=U(x,\mu)U(x+\Hat{\mu},\nu)
   U(x+\Hat{\nu},\mu)^\dagger U(x,\nu)^\dagger,
\label{eq:(2.9)}
\end{align}
and the integer field~$B_{\mu\nu}(x)$ is given by the 't~Hooft fluxes as
\begin{equation}
   B_{\mu\nu}(x)=\begin{cases}
   B_{\mu\nu}&\text{for $x_\mu=L-1$ and~$x_\nu=L-1$},\\
   0&\text{otherwise}.\\
   \end{cases}
\label{eq:(2.10)}
\end{equation}
Note that this field is flat modulo~$N$, $\partial_{[\rho}B_{\mu\nu]}(x)=0\bmod N$,
where $\partial_\rho$ denotes the lattice derivative. The particular form
of~$B_{\mu\nu}(x)$ in~Eq.~\eqref{eq:(2.10)} is not unique and can be changed
by the $\mathbb{Z}_N$ 1-form gauge transformation,
$U(x,\mu)\to e^{2\pi iz_\mu(x)/N}U(x,\mu)$, where $z_\mu(x)\in\mathbb{Z}$. The
invariant characterization of $B_{\mu\nu}(x)$ is the total flux,
$B_{\mu\nu}=\sum_{s,t=0}^{L-1}B_{\mu\nu}(x+s\Hat{\mu}+t\Hat{\nu})\bmod N$. In what
follows, we call the $\mathbb{Z}_N$ 2-form gauge field and/or the 't~Hooft flux
simply, ``$B$-field'' in a moderate abuse of language.

In the next section, we carry out an HMC simulation of the system defined by
the second line of~Eq.~\eqref{eq:(2.8)}, employing configurations of the
$B$-field of the particular form shown in~Eq.~\eqref{eq:(2.10)}. This implies
that we work in a particular gauge of the $\mathbb{Z}_N$ 1-form gauge symmetry;
the observables thus should be invariant under the $\mathbb{Z}_N$ 1-form gauge
transformation, $U(x,\mu)\to e^{2\pi iz_\mu(x)/N}U(x,\mu)$
and~$B_{\mu\nu}(x)\to B_{\mu\nu}(x)+\partial_\mu z_\nu(x)-\partial_\nu z_\mu(x)\bmod N$.

\section{Numerical experiments}
\label{sec:3}
\subsection{``Halfway-updating'' HMC algorithm}
\label{sec:3.1}
In our study, we basically follow the HMC
algorithm~\cite{Duane:1987de},\footnote{Our numerical codes can be found in
\url{https://github.com/o-morikawa/Gaugefields.jl}, which is based on
\texttt{Gaugefields.jl} in the JuliaQCD package~\cite{Nagai:2024yaf}.} having
possible future applications to lattice QCD in mind (see Section~\ref{sec:4}).
However, since the $B$-field takes values in~$\mathbb{Z}$ and has no obvious
conjugate momentum, we have to appropriately modify the HMC algorithm. We see
that the following ``halfway-updating'' HMC algorithm fulfills the detailed
balance, a sufficient condition for the Markov chain Monte Carlo to reproduce
the equilibrium distribution with a given Boltzmann weight.

Let $U$ and~$B$ be the initial configuration of the gauge field and the
$B$-field, respectively. Then,
\begin{enumerate}
\item Generate the initial momentum~$\pi$ being conjugate to~$U$ by the
Gaussian distribution~$P_G(\pi)\sim e^{-\pi^2/2}$.
\item Via the leapfrog method, evolve $\pi$ and~$U$ with respect to the
Hamiltonian~$H(U,\pi,B):=(1/2)\pi^2+S(U,B)$, where the action~$S(U,B)$ is given
by the exponent of the second line of~Eq.~\eqref{eq:(2.8)}, by the MD
time~$\tau/2$. This gives the
mapping~$\{U,\pi\}\stackrel{\tau/2}{\to}\{\Check{U},\Check{\pi}\}$.
\item Update the $B$-field as~$B\to B'$ in a probability~$P_F(B\to B')$. We
assume that $P_F(B\to B')=P_F(B'\to B)$ for any pair $(B,B')$. In our actual
simulations, we set $B_{\mu\nu}'$ for each pair~$(\mu,\nu)$ by a uniform random
number in~$\{0,1,\cdots,N-1\}$.\footnote{%
A technical note on this prescription: Given a $\mathbb{Z}_N$ 2-cochain field
$Z_{\mu\nu}(x)$ on~$T^4$ being flat in the sense that $\prod_cZ_{\mu\nu}(x)=1$,
where the oriented product~$\prod_c$ is taken on 6~faces of a 3D cube~$c$, one
can associate the flux in~Eq.~\eqref{eq:(2.10)} with~$0\leq B_{\mu\nu}<N$ in the
following way. First, define $\Check{B}_{\mu\nu}(x)$
by~$Z_{\mu\nu}(x)=e^{-2\pi i\Check{B}_{\mu\nu}(x)/N}$
and~$0\leq\Check{B}_{\mu\nu}(x)<N$. This $\Check{B}_{\mu\nu}(x)$ is flat but
generally only modulo~$N$, i.e., $\partial_{[\rho}\Check{B}_{\mu\nu]}(x)=0\bmod N$.
On~$T^4$, one can construct a $\mathbb{Z}$ 2-cocycle $M_{\mu\nu}(x)$ such that
$\Bar{B}_{\mu\nu}(x):=\Check{B}_{\mu\nu}(x)-NM_{\mu\nu}(x)$ satisfies
$\partial_{[\rho}\Bar{B}_{\mu\nu]}(x)=0$ ($\Bar{B}_{\mu\nu}(x)$ provides the integral
lift of $H^2(T^4,\mathbb{Z}_N)$ to~$H^2(T^4,\mathbb{Z})$). An explicit method to
construct $M_{\mu\nu}(x)$ on a periodic hypercubic lattice is given
in~Section~4.2 of~Ref.~\cite{Abe:2022nfq}. This $\Bar{B}_{\mu\nu}(x)$ is 1-form
gauge equivalent to~$B_{\mu\nu}(x)$ of the form in~Eq.~\eqref{eq:(2.10)}.
Finally, after the local shift by multiples of~$N$,
$B_{\mu\nu}(x)\to B_{\mu\nu}(x)+NN_{\mu\nu}(x)$, where $N_{\mu\nu}(x)\in\mathbb{Z}$,
one can restrict the range of~$B_{\mu\nu}$ in~Eq.~\eqref{eq:(2.10)}
into~$0\leq B_{\mu\nu}<N$. This argument shows that if the action and observables
are invariant under the 1-form gauge transformation and the local shift of
$B_{\mu\nu}(x)$ by multiples of~$N$, then our algorithm amounts to generating the
$\mathbb{Z}_N$ 2-cochain field $Z_{\mu\nu}(x)$ on~$T^4$.}
\item Again evolve fields as
$\{\Check{U},\Check{\pi}\}\stackrel{\tau/2}{\to}\{U',\pi'\}$ by the MD but now
with respect to the
Hamiltonian~$H(\Check{U},\Check{\pi},B'):=(1/2)\Check{\pi}^2+S(\Check{U},B')$
by the MD time~$\tau/2$.
\item Accept the new configuration $\{U',\pi',B'\}$ under the probability
(the Metropolis test)
\begin{equation}
   P_A(\{U,\pi,B\}\to\{U',\pi',B'\})=\min\left[1,e^{-\Delta H}\right],
\label{eq:(3.1)}
\end{equation}
where $\Delta H:=H(U',\pi',B')-H(U,\pi,B)$.
\item Go back to the first step.
\end{enumerate}
As shown in~Appendix~\ref{sec:A}, this algorithm fulfills the detailed
balance.

The lattice parameters we used are summarized in~Tables~\ref{table:1}
and~\ref{table:2}. In our HMC simulation for the $SU(2)/\mathbb{Z}_2$ theory,
for all lattice parameters, the length of one HMC step~$\tau$
(``one trajectory'') is~$1.0$ in lattice units ($\Delta\tau=0.02$ times $25$ MD
steps for the ``half-way'' and in total there are $50$ MD steps for one HMC
step). For our largest and finest lattice ($\beta=2.6$ and~$L=20$), the
Metropolis acceptance was~$\sim86\%$. We also carry out the HMC simulation for
the $SU(2)$ theory using the conventional HMC algorithm. For this also, for all
lattice parameters, $\tau=1.0$ in lattice units ($\Delta\tau=0.02$ and there
are $50$ MD steps).
\begin{table}[htbp]
\caption{The lattice parameters in our HMC simulation in the
$SU(2)/\mathbb{Z}_2$ theory.}
\label{table:1}
\begin{center}
\begin{tabular}{crllc}
\toprule
$\beta$&$L$&$a\sqrt{\sigma}$&$La\sqrt{\sigma}$&$N$ configs.\\
\midrule
$2.4$&$8$&$0.2673$&$2.138$&$5000$\\
$2.5$&$12$&$0.186$&$2.23$&$5000$\\
$2.6$&$16$&$0.1326$&$2.122$&$1744$\\
\midrule
$2.4$&$10$&$0.2673$&$2.673$&$5000$\\
$2.5$&$14$&$0.186$&$2.60$&$5387$\\
$2.6$&$20$&$0.1326$&$2.652$&$1046$\\
\bottomrule
\end{tabular}
\end{center}
\end{table}
\begin{table}[htbp]
\caption{The lattice parameters in our HMC simulation in the
$SU(2)$ theory.}
\label{table:2}
\begin{center}
\begin{tabular}{crllc}
\toprule
$\beta$&$L$&$a\sqrt{\sigma}$&$La\sqrt{\sigma}$&$N$ configs.\\
\midrule
$2.4$&$8$&$0.2673$&$2.138$&$5000$\\
$2.5$&$12$&$0.186$&$2.23$&$5000$\\
$2.6$&$16$&$0.1326$&$2.122$&$1635$\\
\bottomrule
\end{tabular}
\end{center}
\end{table}

For the $SU(2)$ theory (i.e., without the $B$-field), we referred to the
mapping between $\beta$ and~$a\sqrt{\sigma}$, where $\sigma$ is the string
tension, given in~Ref.~\cite{Teper:1998kw}. We used this mapping also for the
$SU(2)/\mathbb{Z}_2$ theory (i.e., with the $B$-field). This point might be
subtle, however, because the conventional Wilson line operator is not gauge
invariant in the $SU(2)/\mathbb{Z}_2$ theory. A more satisfactory way would be
to use the gradient flow~\cite{Luscher:2010iy} such that the value
of~$a/\sqrt{t}$, where $t/a^2$ is the flow time in lattice units, at which the
expectation value of the local
operator~\eqref{eq:(3.2)} $t^2\langle E(t)\rangle$ becomes (say) $0.3$. In any
case, in the $SU(2)/\mathbb{Z}_2$ theory, the dependence of the integrated
autocorrelation time on the lattice spacing is quite weak and this subtlety
should not crucially change our conclusion.

For each of lattice parameters, we stored configurations per each 10 units of
MD time (i.e., per every 10 trajectories).

\subsection{Autocorrelation functions}
\label{sec:3.2}
We are primarily interested in the HMC history of the topological charge~$Q$.
It is possible to construct a geometrical definition of the lattice topological
charge~$Q$ in the $SU(N)$ theory with the $B$-field, which exactly
ensures fractional values~\eqref{eq:(1.2)}, by combining the seminal idea
in~Ref.~\cite{Luscher:1981zq} and the $\mathbb{Z}_N$ 1-form gauge
invariance~\cite{Abe:2023ncy}. Its actual implementation, however, appears quite
complicated. Here, therefore, we instead employ the following definition of~$Q$
on the basis of the gradient flow~\cite{Luscher:2010iy}.

We adopt the tree-level improved linear combination of the clover and
rectangular definitions given in~Ref.~\cite{Alexandrou:2015yba}, by including
the $B$-field in each of the plaquettes for the 1-form gauge invariance. We
substitute a field configuration smeared by the gradient flow in this
definition. Here, the gradient flow refers to the lattice action in the second
line of~Eq.~\eqref{eq:(2.8)} and thus the flow equation itself depends also on
the $B$-field.\footnote{We do not consider the ``flow'' of the $B$-field
itself. We admit that the renormalizability~\cite{Luscher:2011bx,Hieda:2016xpq}
of this gradient flow with a fixed $B$-field is an open problem, although we
expect it because the local dynamics should be insensitive to whether we
simulate $SU(N)$ or $SU(N)/\mathbb{Z}_N$.} Therefore, the gradient flow acts so
that the modified plaquette variables~$e^{2\pi iB_{\mu\nu}}P(x,\mu,\nu)$,
\emph{not\/} $P(x,\mu,\nu)$, become smooth.\footnote{In other words, the
smoothness on the lattice, the
admissibility~\cite{Luscher:1981zq,Hernandez:1998et}, is replaced
by~$\|\bm{1}-e^{2\pi iB_{\mu\nu}}P(x,\mu,\nu)\|<\varepsilon$ for a certain small
constant~$\varepsilon$~\cite{Abe:2023ncy}.} Since the $B$-field is not
differentiable (it takes values in~$\mathbb{Z}$), this difference in the
meaning of the smoothness has a drastic effect which can make the topological
charge fractional.\footnote{We would like to thank Yuya Tanizaki for originally
sharing this idea with us.} In the present exploratory study, we measure the
topological charge of flowed configurations only at a flow time $t=(0.7L)^2/8$
in lattice units, where $L$ is the lattice size, corresponding to the smeared
or diffused length~$\sim\sqrt{8t}=0.7L$; in this paper, we do not study how the
results depend on this choice of the flow time.

\begin{figure}[htbp]
\centering
\includegraphics[width=8cm]{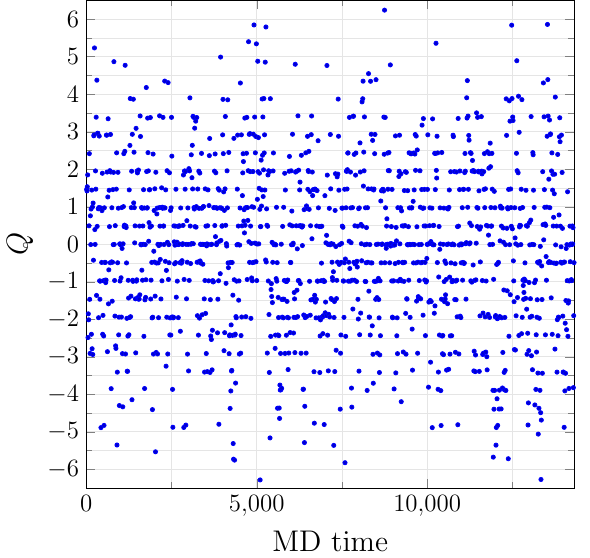}
\caption{The HMC history of the topological charge~$Q$ in the
$SU(2)/\mathbb{Z}_2$ theory. $\beta=2.6$ and~$L=16$.}
\label{fig:1}
\end{figure}

In Fig.~\ref{fig:1}, we depict the HMC history of the topological charge~$Q$ in
the $SU(2)/\mathbb{Z}_2$ theory ($\beta=2.6$ and~$L=16$). We clearly observe
that $Q$ takes (approximately) fractional values as expected in the present
$SU(2)/\mathbb{Z}_2$ case, $Q=1/2+\mathbb{Z}$. Also, the distribution of~$Q$
spreads in a wide range without any obvious autocorrelation in the HMC history;
it appears that the topological sectors are shuffled by the dynamical effect of
the $B$-field. 

\begin{figure}[htbp]
\centering
\includegraphics[width=8cm]{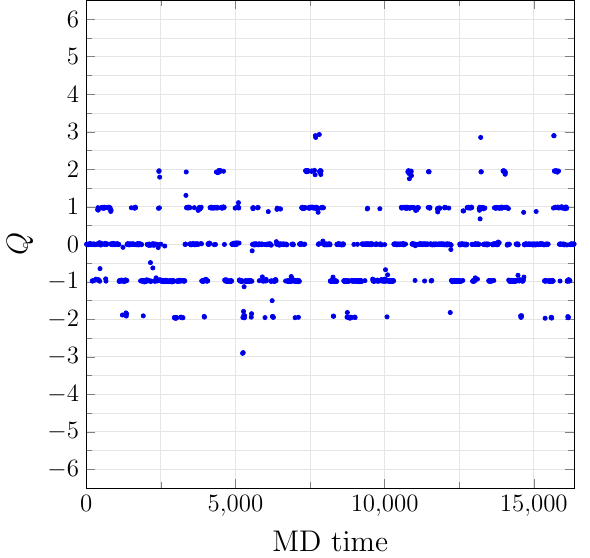}
\caption{The HMC history of the topological charge~$Q$ in the
$SU(2)$ theory (i.e., without the $B$-field). $\beta=2.6$ and~$L=16$. The
topological freezing is clearly seen.}
\label{fig:2}
\end{figure}

For a comparison, in~Fig.~\ref{fig:2}, we depict the HMC history of the
topological charge~$Q$ in the HMC simulation of the $SU(2)$ theory. In
contrast, we clearly observe the tendency that $Q$ is stuck and the topological
freezing of $O(500)$ MD time.

The HMC history of~$Q$ for other lattice parameters in conjunction with the
histogram of~$Q$ are presented in~Appendix~\ref{sec:B}.

\begin{figure}[htbp]
\centering
\includegraphics[width=8cm]{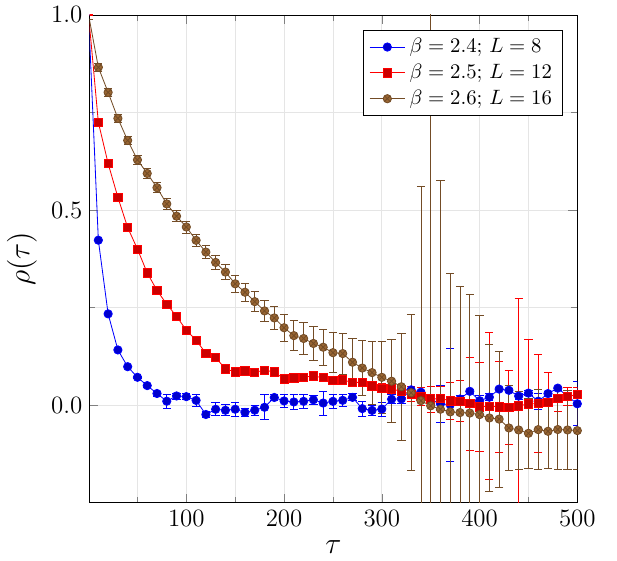}
\caption{The normalized autocorrelation functions of the topological charge~$Q$
in the conventional HMC simulation of the $SU(2)$ theory. The horizontal axis
is the MD time. The cases in~Table~\ref{table:2} with approximately identical
physical lattice sizes are plotted.}
\label{fig:3}
\end{figure}

We expect that the autocorrelation length increases as the system approaches
the critical point (i.e., the continuum limit). In~Fig.~\ref{fig:3}, we plot
the normalized autocorrelation functions of the topological charge~$Q$ in the
$SU(2)$ theory as a function of the MD time; we follow the definition of the
autocorrelation function in~Appendix~E of~Ref.~\cite{Luscher:2004pav}.

\begin{figure}[htbp]
\centering
\includegraphics[width=8cm]{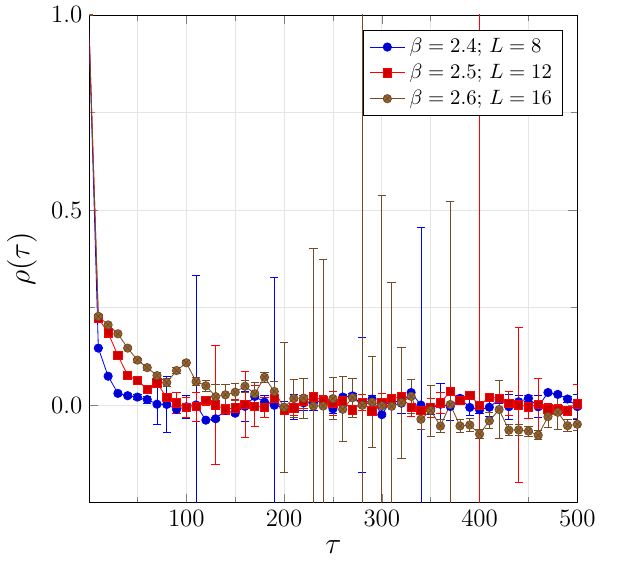}
\caption{The normalized autocorrelation functions of the topological charge~$Q$
in the HMC simulation of the $SU(2)/\mathbb{Z}_2$ theory (i.e., with the
$B$-field). The horizontal axis is the MD time. The cases of first three rows
in~Table~\ref{table:1} with approximately identical physical lattice sizes are
plotted. The lattice parameters are identical to those in~Fig.~\ref{fig:3}.}
\label{fig:4}
\end{figure}

\begin{figure}[htbp]
\centering
\includegraphics[width=8cm]{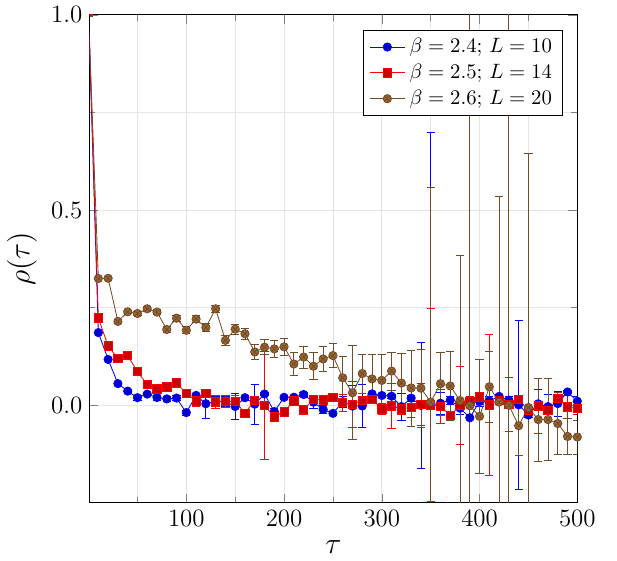}
\caption{The normalized autocorrelation functions of the topological charge~$Q$
in the HMC simulation of the $SU(2)/\mathbb{Z}_2$ theory. The horizontal axis
is the MD time. The cases of latter three rows of~Table~\ref{table:1} of
approximately identical physical lattice sizes are plotted.}
\label{fig:5}
\end{figure}

Figure~\ref{fig:3} is for the $SU(2)$ theory without the $B$-field.
In~Fig.~\ref{fig:4}, we plot the normalized autocorrelation function of~$Q$
in the $SU(2)/\mathbb{Z}_2$ theory with the dynamical $B$-field. We observe
that the autocorrelation is drastically reduced.
Figure~\ref{fig:5} shows the normalized autocorrelation function of the
topological charge for the latter three rows of~Table~\ref{table:1}.

In~Fig.~\ref{fig:6}, we plot the integrated autocorrelation lengths (we follow
the definition in~Appendix~E of~Ref.~\cite{Luscher:2004pav}) of the
topological charge~$Q$ and it square~$Q^2$ as a function of the lattice
spacing~$a$ in units of the string tension~$\sigma$.
\begin{figure}[htbp]
\centering
\includegraphics[width=8cm]{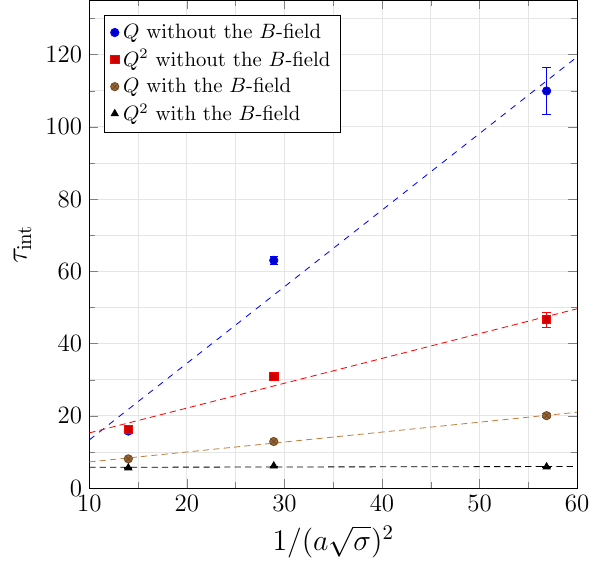}
\caption{The integrated autocorrelation lengths in lattice units of the
topological charge~$Q$ and of the square~$Q^2$ as the function of the lattice
spacing. For the cases in the first three rows of~Table~\ref{table:1} and
Table~\ref{table:2}.}
\label{fig:6}
\end{figure}
The general consideration~\cite{Luscher:2011kk} indicates that the integrated
autocorrelation length in the HMC algorithm increases as~$a^{-2}$. We fairly
well observe this scaling behavior in~Fig.~\ref{fig:6}. However, the slope is
quite small for the $SU(2)/\mathbb{Z}_2$ theory with our HMC algorithm.

\begin{figure}[htbp]
\centering
\includegraphics[width=8cm]{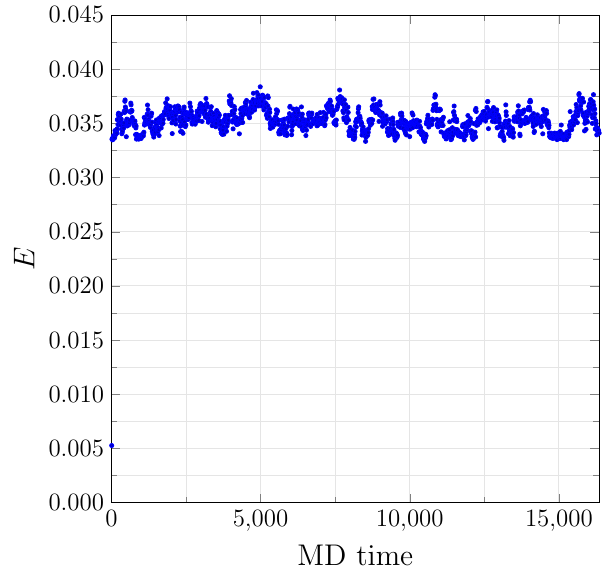}
\caption{The HMC history of~$E(t)$~\eqref{eq:(3.2)} in the $SU(2)$ theory
(i.e., without the $B$-field). $\beta=2.6$ and~$L=16$.}
\label{fig:7}
\end{figure}

\begin{figure}[htbp]
\centering
\includegraphics[width=8cm]{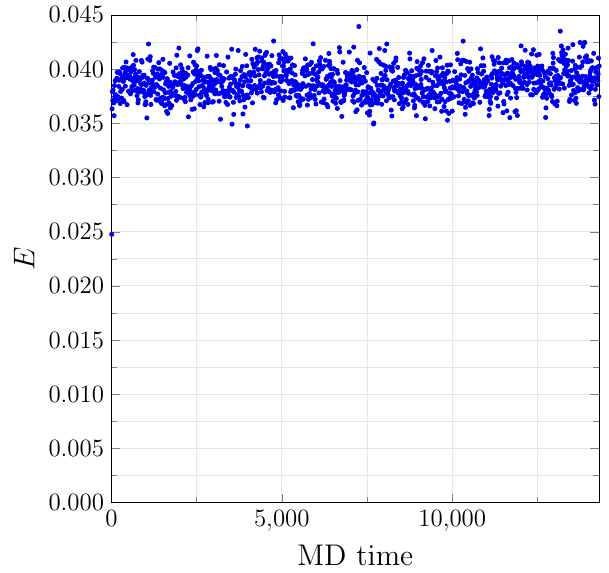}
\caption{The HMC history of~$E(t)$~\eqref{eq:(3.2)} in the $SU(2)/\mathbb{Z}_2$
theory (i.e., with the $B$-field). $\beta=2.6$ and~$L=16$.}
\label{fig:8}
\end{figure}

The drastic reduction of the autocorrelation in~$Q$ and in~$Q^2$ is, although
quite impressive, somehow expected because the random choice of the $B$-field
inevitably shuffles the topological charge. This does not, however, necessarily
imply that the autocorrelation of any observables is reduced in the
$SU(2)/\mathbb{Z}_2$ theory. To examine this point, we measure the HMC history
of another observable which is expected to have a large overlap with slow modes
of the HMC algorithm. It is given by the so-called
``energy-operator''~\cite{Luscher:2011kk} (we implicitly assume the average
over~$x\in\Gamma$),
\begin{equation}
   E(t):=\frac{1}{2}\tr\left[F_{\mu\nu}(t,x)F_{\mu\nu}(t,x)\right],
\label{eq:(3.2)}
\end{equation}
where the field strength on the right-hand side is given by the gradient flow
at the flow time~$t$~\cite{Luscher:2010iy} (in the $SU(2)/\mathbb{Z}_2$ theory,
we multiply each of plaquettes by the $B$-field). The flow time is fixed
to~$t=(0.7L)^2/8$. In~Figs.~\ref{fig:7} and~\ref{fig:8}, we plot the HMC
histories of~$E(t)$ for the $SU(2)$ theory and for the $SU(2)/\mathbb{Z}_2$
theory, respectively ($\beta=2.6$ and~$L=16$).
In Fig.~\ref{fig:7}, we clearly observe the large autocorrelation of $O(100)$
MD time, whereas in the $SU(2)/\mathbb{Z}_2$ theory of~Fig.~\ref{fig:8}, we do
not see any notable autocorrelation; in fact, the integrated autocorrelation
time is of~$O(20)$ MD time in the latter. These observations strongly indicate
that our HMC algorithm for the $SU(2)/\mathbb{Z}_2$ or more generally the
$SU(N)/\mathbb{Z}_N$ theory reduces the autocorrelation in general physical
quantities drastically.

\begin{figure}[htbp]
\centering
\includegraphics[width=8cm]{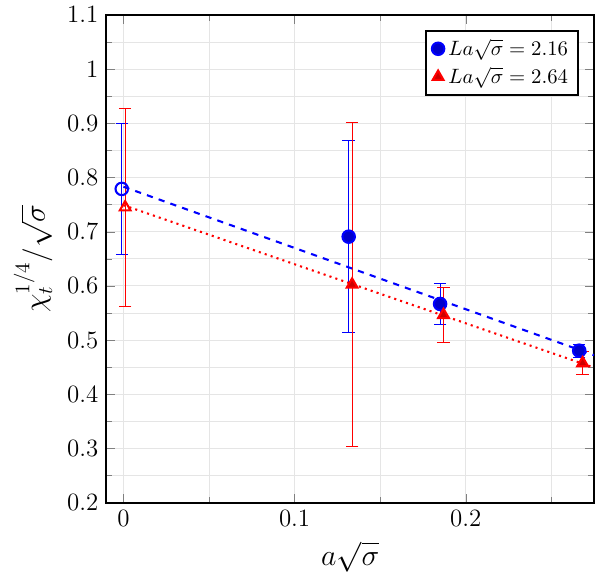}
\caption{The continuum extrapolations of the topological
susceptibility~$\chi_t$~\eqref{eq:(3.3)} in the $SU(2)/\mathbb{Z}_2$ theory.
The two lines correspond to two sequences corresponding to two different
physical volumes; the first/blue line corresponds to the first three lattice
parameters in~Table~\ref{table:1} whereas the second/red line corresponds to
the latter three lattice parameters in~Table~\ref{table:1}.}
\label{fig:9}
\end{figure}

\begin{figure}[htbp]
\centering
\includegraphics[width=8cm]{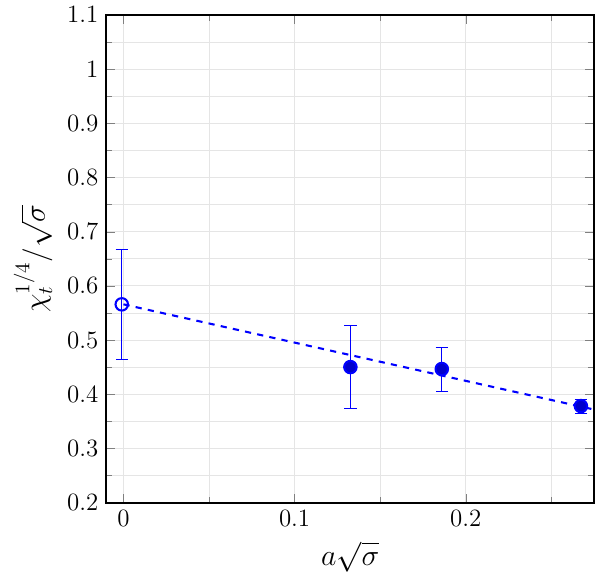}
\caption{The continuum extrapolation of the topological
susceptibility~$\chi_t$~\eqref{eq:(3.3)} in the $SU(2)$ theory. The result is
consistent with the one in~Ref.~\cite{Teper:1998kw},
$\chi_t^{1/4}/\sqrt{\sigma}=0.486(10)$.}
\label{fig:10}
\end{figure}

Finally, to have some idea how large is the finite size effect which provides
the difference between $SU(2)/\mathbb{Z}_2$ and~$SU(2)$ on local observables,
in Figs.~\ref{fig:9} and~\ref{fig:10}, we plot the topological susceptibility
\begin{equation}
   \chi_t:=\frac{1}{(La)^4}\left\langle Q^2\right\rangle
\label{eq:(3.3)}
\end{equation}
in units of the string tension~$\sigma$. For all lattice parameters, the first
$50$ configurations ($500$ MD time) are omitted for thermalization and
statistical errors are estimated by the jackknife method with a bin size~$40$,
corresponding to $40$ configurations and $400$ units of MD time. Although the
error bars for the $SU(2)/\mathbb{Z}_2$ and $SU(2)$ theories in
Figs.~\ref{fig:9} and~\ref{fig:10} are of almost the same order, we observed
that the jackknife errors in the $SU(2)/\mathbb{Z}_2$ cases are much more
quickly saturated compared to the $SU(2)$ cases, as expected from the shortness
of the autocorrelation.

\begin{figure}[htbp]
\centering
\includegraphics[width=8cm]{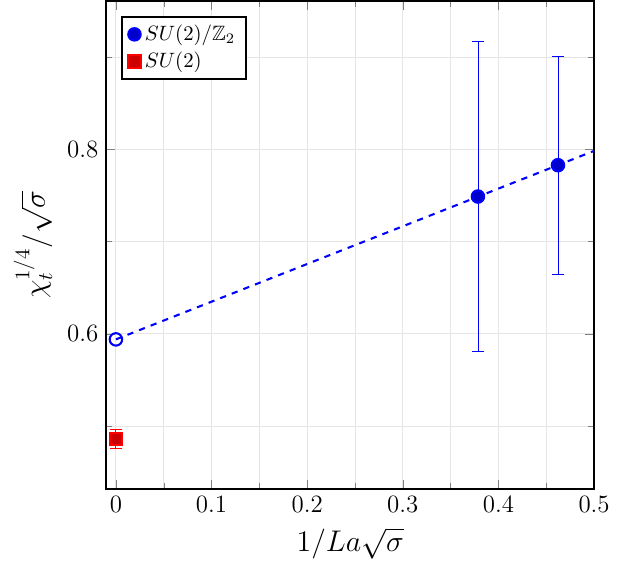}
\caption{Continuum extrapolations
in~Fig.~\ref{fig:9} as a function of the physical volume. A naive linear
extrapolation of central values to the infinite volume appears consistent with
the result in the $SU(2)$ theory, indicated by~$SU(2)$ in the figure,
$\chi_t^{1/4}/\sqrt{\sigma}=0.486(10)$~\cite{Teper:1998kw}, considering large
statistical error bars in our results.}
\label{fig:11}
\end{figure}

In Fig.~\ref{fig:11}, we combined the two continuum extrapolations
in~Fig.~\ref{fig:9} as a function of the physical lattice size. A naive
linear extrapolation of central values to the infinite volume appears
consistent with the result in the $SU(2)$ theory~\cite{Teper:1998kw},
considering large statistical error bars in our results. To conclude the
validity of the present approach to the $SU(2)$ theory from the
$SU(2)/\mathbb{Z}_2$ theory, however, we need further statistics for larger and
finer lattices.

\section{Incorporation of fundamental quarks}
\label{sec:4}
One clearly wants to incorporate matter fields in the present framework. The
incorporation of the adjoint matter fields such as the gaugino in the
$\mathcal{N}=1$ supersymmetric Yang--Mills theory would be straightforward at
least in principle because it is blind to the center part of the gauge
group~$SU(N)$, $\mathbb{Z}_N$; thus $SU(N)/\mathbb{Z}_N$ can be gauged. A
better control on errors in the topological susceptibility could be quite
useful for finding a supersymmetric point of the bare lattice gluino mass
parameter~\cite{Curci:1986sm}, since a massless fermion implies the vanishing
of the topological susceptibility.

It is, however, not obvious whether it is possible to incorporate ``quarks,''
fermions in the fundamental representation of~$SU(N)$ and this point can be an
obstacle to extending the present framework to lattice QCD. In what follows, we
consider a possible method to avoid this obstruction.

First, when the $B$-field is fixed, i.e., when it is \emph{not dynamical}, one
may incorporate quarks as follows~\cite{Tanizaki:2022ngt}: We first define
boundary conditions of the fundamental fermions by
\begin{align}
   \Tilde\psi(x+L\Hat{\mu})
   &=e^{-i\alpha_\mu(x)/N}g_\mu(x)^\dagger\Tilde{\psi}(x),
\notag\\
   \Tilde{\Bar{\psi}}(x+L\Hat{\mu})
   &=\Tilde{\Bar{\psi}}(x)g_\mu(x)e^{i\alpha_\mu(x)/N},
\label{eq:(4.1)}
\end{align}
where $x_\mu=0$. Here, transition functions~$g_\mu(x)$ are identified with those
in~Eq.~\eqref{eq:(2.1)} and $e^{-i\alpha_\mu(x)/N}$ is the transition functions of
the baryon number~$U(1)$ ($U(1)_B$) principal bundle; the factor~$1/N$ arises
since we set the quark $U(1)_B$ charge~$1/N$. Then, postulating the cocycle
condition at~$x_\mu=x_\nu=0$,
\begin{equation}
   e^{i\alpha_\mu(x)/N}e^{i\alpha_\nu(x+L\Hat{\mu})/N}
   e^{-i\alpha_\mu(x+L\Hat{\nu})/N}e^{-i\alpha_\nu(x)/N}
   =e^{-2\pi iB_{\mu\nu}/N},
\label{eq:(4.2)}
\end{equation}
the $\mathbb{Z}_N$ factors in this relation cancel the $\mathbb{Z}_N$ factors
in~Eq.~\eqref{eq:(2.2)}; the fermion fields can thus be consistently defined
on~$T^4$ although it belongs to the fundamental representation
of~$SU(N)$~\cite{Tanizaki:2022ngt}. Corresponding to~Eq.~\eqref{eq:(4.1)}, the
$U(1)_B$ gauge potential 1-form~$A_B$ in the charge~$1/N$ representation
possess the boundary conditions
\begin{equation}
   A_B(x+L\Hat{\mu})=A_B(x)+\frac{1}{N}\mathrm{d}\alpha_\mu(x),
\label{eq:(4.3)}
\end{equation}
where $x_\mu=0$. This $U(1)_B$ bundle is characterized by the first Chern
numbers,
\begin{equation}
   \int_{\text{$\mu\nu$ plane}}\mathrm{d}A_B
   =2\pi\left(\frac{B_{\mu\nu}}{N}+\mathbb{Z}\right),
\label{eq:(4.4)}
\end{equation}
where we have used Eqs.~\eqref{eq:(4.3)} and~\eqref{eq:(4.2)}. Note that we
can introduce mass terms for quarks in this setup as far as they preserve
the $U(1)_B$ symmetry, not necessarily the
flavor~$SU(N_f)$~\cite{Tanizaki:2022ngt}. Thus, this system is quite different
from the $\mathbb{Z}_N$ QCD~\cite{Kouno:2012zz,Iritani:2015ara} with the
$SU(N)$ flavor symmetry.

In lattice gauge theory, we introduce the $U(1)_B$ link variables as
$\Tilde{U}_B(x,\mu)\sim\exp(i\int_x^{x+\Hat{\mu}}A_B)$. Corresponding
to~Eq.~\eqref{eq:(4.3)}, $\Tilde{U}_B(x,\mu)$ obey boundary conditions
\begin{equation}
   \Tilde{U}_B(x+L\Hat{\mu},\nu)
   =e^{-i\alpha_\mu(x)/N}\Tilde{U}_B(x,\nu)e^{i\alpha_\mu(x+\Hat{\nu})/N}.
\label{eq:(4.5)}
\end{equation}
Under the change of link variables analogous to~Eq.~\eqref{eq:(2.7)},
\begin{equation}
   \Tilde{U}_B(x,\mu)=
   \begin{cases}
   U_B(x,\mu)e^{i\alpha_\mu(x)/N}&\text{for $x_\mu=L-1$},\\   
   U_B(x,\mu)&\text{otherwise},\\
   \end{cases}
\label{eq:(4.6)}
\end{equation}
we find that
\begin{equation}
   \Tilde{P}_B(x,\mu,\mu)
   =e^{2\pi iB_{\mu\nu}(x)/N}P_B(x,\mu,\nu),
\label{eq:(4.7)}
\end{equation}
where
\begin{align}
   \Tilde{P}_B(x,\mu,\nu)
   &:=\Tilde{U}_B(x,\mu)\Tilde{U}_B(x+\Hat{\mu},\nu)
   \Tilde{U}_B(x+\Hat{\nu},\mu)^*\Tilde{U}_B(x,\nu)^*,
\notag\\
   P_B(x,\mu,\nu)
   &:=U_B(x,\mu)U_B(x+\Hat{\mu},\nu)U_B(x+\Hat{\nu},\mu)^*U_B(x,\nu)^*,
\label{eq:(4.8)}
\end{align}
and $B_{\mu\nu}(x)$ is given by~Eq.~\eqref{eq:(2.10)}. The variables $U_B$ are
regarded as obeying periodic boundary conditions,
$U_B(x+L\Hat{\mu},\nu)=U_B(x,\nu)$, where $x_\mu=0$. Using the boundary
conditions, Eqs.~\eqref{eq:(4.1)}, \eqref{eq:(4.5)} and~\eqref{eq:(2.3)}, and
then the change of variables, Eqs.~\eqref{eq:(2.7)} and~\eqref{eq:(4.6)}, we
find that lattice hopping terms of the fundamental fermion take the following
forms,
\begin{align}
   \Tilde{\Bar{\psi}}(x)\Tilde{U}_B(x,\mu)\Tilde{U}(x,\mu)
   \Tilde{\psi}(x+\Hat{\mu})
   &=\Bar{\psi}(x)U_B(x,\mu)U(x,\mu)\psi(x+\Hat{\mu}),
\notag\\
   \Tilde{\Bar{\psi}}(x)\Tilde{U}(x-\Hat{\mu},\mu)^\dagger
   \Tilde{U}_B(x-\Hat{\mu},\mu)^*
   \Tilde{\psi}(x-\Hat{\mu})
   &=\Bar{\psi}(x)U(x-\Hat{\mu},\mu)^\dagger U_B(x-\Hat{\mu},\mu)^*
   \psi(x-\Hat{\mu}),
\label{eq:(4.9)}
\end{align}
\emph{everywhere\/} on the lattice~$\Gamma$ including boundaries. In this
expression, fermion variables $\psi(x)$ and~$\Bar{\psi}(x)$ are understood to
obey the periodic boundary conditions, $\psi(x+L\Hat{\mu})=\psi(x)$
and~$\Bar{\psi}(x+L\Hat{\mu})=\Bar{\psi}(x)$ for~$x_\mu=0$.

So far, the $U(1)_B$ gauge field~$\Tilde{U}_B(x,\mu)$ is regarded as a
non-dynamical background. However, since the $B$-field is dynamical in the
$SU(N)/\mathbb{Z}_N$ theory and $\Tilde{U}_B(x,\mu)$ depends on the $B$-field
through the boundary conditions~\eqref{eq:(4.5)}, $\Tilde{U}_B(x,\mu)$
inevitably becomes dynamical in the $SU(N)/\mathbb{Z}_N$ theory.

The Wilson plaquette action for the $U(1)_B$
gauge field would be
\begin{align}
   &\exp\left\{
   \beta_B\sum_{x\in\Gamma}\sum_{\mu<\nu}
   \Re\left[\Tilde{P}_B(x,\mu,\nu)-1\right]
   \right\}
\notag\\
   &=
   \exp\left\{
   \beta_B\sum_{x\in\Gamma}\sum_{\mu<\nu}
   \Re\left[e^{2\pi iB_{\mu\nu}(x)/N}P_B(x,\mu,\nu)-1\right]
   \right\},
\label{eq:(4.10)}
\end{align}
where $\beta_B$ is the bare coupling. However, of course, the dynamical $U(1)_B$
gauge field is unwanted for the sake to simulate QCD. For a possible solution
on this point, we make the $U(1)_B$ gauge boson super-heavy by the
St\"uckelberg mechanism on the lattice. That is, we introduce a $U(1)$-valued
dynamical scalar field~$\Omega(x)\in U(1)$ and add the lattice action
\begin{align}
   &S_{\text{St\"uckelberg}}
\notag\\
   &:=\frac{1}{2}\sum_{x\in\Gamma}\sum_\mu
   \left[1-\Tilde{\Omega}(x+\hat{\mu})^*
   \Tilde{U}_B(x,\mu)^{N*}\Tilde{\Omega}(x)\right]
   \left[\Tilde{\Omega}(x)^*\Tilde{U}_B(x,\mu)^N
   \Tilde{\Omega}(x+\hat{\mu})-1\right]
\notag\\
   &=\frac{1}{2}\sum_{x\in\Gamma}\sum_\mu
   \left[1-\Omega(x+\hat{\mu})^*U_B(x,\mu)^{N*}\Omega(x)\right]
   \left[\Omega(x)^*U_B(x,\mu)^N\Omega(x+\hat{\mu})-1\right].
\label{eq:(4.11)}
\end{align}
In this expression, $\Tilde{\Omega}(x)$ obeys the twisted boundary conditions,
$\Tilde{\Omega}(x+L\Hat{\mu})=e^{-i\alpha_\mu(x)}\Tilde{\Omega}(x)$ ($x_\mu=0$),
while $\Omega(x)$ obeys the periodic boundary conditions,
$\Omega(x+L\Hat{\mu})=\Omega(x)$. $S_{\text{St\"uckelberg}}$ is invariant under the
$U(1)_B$ 0-form and $\mathbb{Z}_N$ 1-form gauge transformations. The gauge
fixing of~$U(1)_B$ of the form $\Omega(x)=1$ then shows that the $U(1)_B$ gauge
boson acquires the mass of~$O(a^{-2})$ and we expect that this freedom decouples
in the continuum limit.

\section{Conclusion}
\label{sec:5}
In this paper, we carried out an HMC simulation of the $SU(2)/\mathbb{Z}_2$
Yang--Mills theory, in which the $\mathbb{Z}_N$ 2-form flat gauge field (the
't~Hooft flux) is treated as a dynamical variable. We observed that our HMC
algorithm in the $SU(2)/\mathbb{Z}_2$ theory drastically reduces the
autocorrelation lengths of the topological charge and of the
``energy-operator'' defined by the gradient flow. We thus infer that, provided
that sufficiently large lattice volumes are available, the HMC simulation of
the $SU(N)/\mathbb{Z}_N$ theory could be employed as an alternative for the
simulation of the $SU(N)$ Yang--Mills theory with a very efficient sampling of
topological sectors. Toward applications in lattice QCD, we also presented a
possible method to incorporate quarks. Further tests on these ideas are to be
awaited.

\section*{Acknowledgments}
We would like to thank Yui Hayashi, Yuki Nagai, Yuya Tanizaki, Akio Tomiya, and
Hiromasa Watanabe for helpful discussions.
We appreciate the opportunity of the
discussion during the YITP--RIKEN iTHEMS conference ``Generalized symmetries in
QFT 2024'' (YITP-W-24-15) in execution of this work.
Numerical computations in this paper were carried out on Genkai, a
supercomputer system of the Research Institute for Information Technology
(RIIT), Kyushu University.
The work of M.A.\ was supported by Kyushu University Innovator Fellowship
Program in Quantum Science Area.
O.M.\ acknowledges the RIKEN Special Postdoctoral Researcher Program.
The work of H.S. was partially supported by Japan Society for the Promotion of
Science (JSPS) Grant-in-Aid for Scientific Research, JP23K03418.

\appendix

\section{Halfway HMC fulfills the detailed balance}
\label{sec:A}
In this appendix, we give a proof that the halfway HMC algorithm
in~Section~\ref{sec:3.1} fulfills the detailed balance, a sufficient condition
for the Markov chain Monte Carlo to reproduce the equilibrium distribution.

Since $B$ and~$B'$ are fixed within the MD time from~$0$ to~$\tau/2$
and from~$\tau/2$ to~$\tau$, respectively, the leapfrog integrator possesses
the invertibility, i.e., if there exists an MD trajectory such that
$\{U,\pi\}_B\stackrel{\tau/2}{\to}\{\Check{U},\Check{\pi}\}_B$, then
$\{\Check{U},-\Check{\pi}\}_B\stackrel{\tau/2}{\to}\{U,-\pi\}_B$ holds (if
there exists an MD trajectory such that
$\{\Check{U},\Check{\pi}\}_{B'}\stackrel{\tau/2}{\to}\{U',\pi'\}_{B'}$, then
$\{U',-\pi'\}_{B'}\stackrel{\tau/2}{\to}\{\Check{U},-\Check{\pi}\}_{B'}$ holds).
That is, probabilities associated with the MD satisfy
\begin{align}
   P_M(\{U,\pi\}\stackrel{\tau/2}{\to}\{\Check{U},\Check{\pi}\})|_B
   &=
   P_M(\{\Check{U},-\Check{\pi}\}\stackrel{\tau/2}{\to}\{U,-\pi\})|_B,
\notag\\
   P_M(\{\Check{U},\Check{\pi}\}\stackrel{\tau/2}{\to}\{U',\pi'\})|_{B'}
   &=
   P_M(\{U',-\pi'\}\stackrel{\tau/2}{\to}\{\Check{U},-\Check{\pi}\})|_{B'},
\label{eq:(A1)}
\end{align}
We also know that the Metropolis test probability in~Eq.~\eqref{eq:(3.1)}
satisfies
\begin{equation}
   e^{-H(U,\pi,B)}P_A(\{U,\pi,B\}\to\{U',\pi',B'\})
   =e^{-H(U',\pi',B')}P_A(\{U',\pi',B'\}\to\{U,\pi,B\}).
\label{eq:(A2)}
\end{equation}
We note that the total probability of the present halfway HMC is given by
\begin{align}
   &P(\{U,B\}\to\{U',B'\})
\notag\\
   &=\int d\pi\,d\pi'\,P_G(\pi)
   P_M(\{U,\pi\}\stackrel{\tau/2}{\to}\{\Check{U},\Check{\pi}\})|_BP_F(B\to B')
   P_M(\{\Check{U},\Check{\pi}\}\stackrel{\tau/2}{\to}\{U',\pi'\})|_{B'}
\notag\\
   &\qquad{}
   \times P_A(\{U,\pi,B\}\to\{U',\pi',B'\}).
\label{eq:(A3)}
\end{align}
From this, noting $H(U,\pi,B)=(1/2)\pi^2+S(U,B)$ and~Eq.~\eqref{eq:(A2)}, we
find
\begin{align}
   &e^{-S(U,B)}P(\{U,B\}\to\{U',B'\})
\notag\\
   &=\int d\pi\,d\pi'\,e^{-H(U',\pi',B')}
\notag\\
   &\qquad{}
   \times
   P_M(\{U,\pi\}\stackrel{\tau/2}{\to}\{\Check{U},\Check{\pi}\})|_B
   P_F(B\to B')
   P_M(\{\Check{U},\Check{\pi}\}\stackrel{\tau/2}{\to}\{U',\pi'\})|_{B'}
\notag\\
   &\qquad\qquad{}
   \times
   P_A(\{U',\pi',B'\}\to\{U,\pi,B\}).
\label{eq:(A4)}
\end{align}
Then, noting $H(U',\pi',B')=(1/2)\pi^{\prime2}+S(U',B')$ and Eq.~\eqref{eq:(A1)},
\begin{align}
   &e^{-S(U,B)}P(\{U,B\}\to\{U',B'\})
\notag\\
   &=e^{-S(U',B')}\int d\pi\,d\pi'\,P_G(\pi')
\notag\\
   &\qquad{}
   \times
   P_M(\{U',-\pi'\}\stackrel{\tau/2}{\to}\{\Check{U},-\Check{\pi}\})|_{B'}
   P_F(B'\to B)
   P_M(\{\Check{U},-\Check{\pi}\}\stackrel{\tau/2}{\to}\{U,-\pi\})|_B
\notag\\
   &\qquad\qquad{}
   \times
   P_A(\{U',\pi',B'\}\to\{U,\pi,B\})
\notag\\
   &=e^{-S(U',B')}P(\{U',B'\}\to\{U,B\}),
\label{eq:(A5)}
\end{align}
where we have used $P_F(B\to B')=P_F(B'\to B)$. This is the detailed balance.

\section{The HMC history and the histogram of~$Q$.}
\label{sec:B}
\begin{figure}[htbp]
\centering
\begin{subfigure}{0.45\columnwidth}
\centering
\includegraphics[width=\columnwidth]{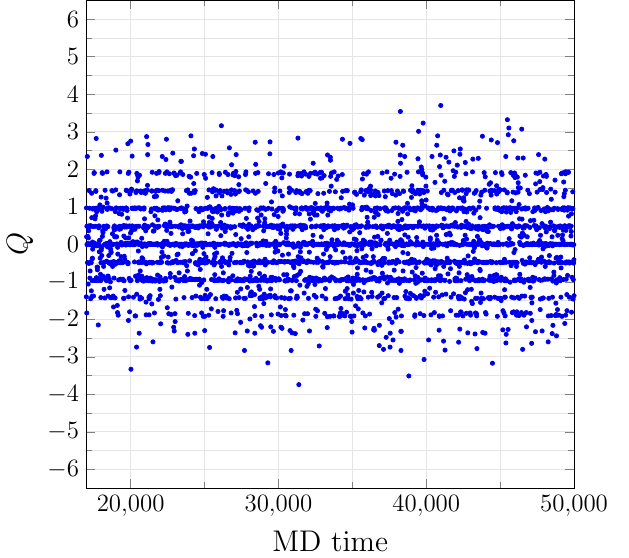}
\caption{$\beta=2.4$ and~$L=8$.}
\label{fig:B1a}
\end{subfigure}
\hspace{6mm}
\begin{subfigure}{0.45\columnwidth}
\centering
\includegraphics[width=\columnwidth]{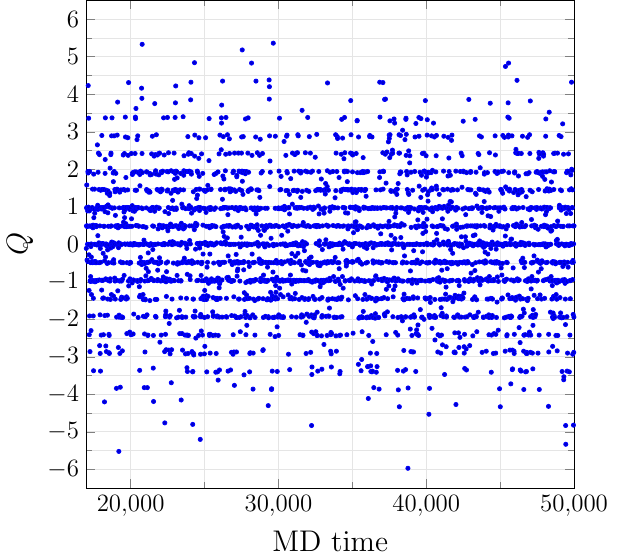}
\caption{$\beta=2.5$ and~$L=12$.}
\label{fig:B1b}
\end{subfigure}
\\[8mm]
\begin{subfigure}{0.45\columnwidth}
\centering
\includegraphics[width=\columnwidth]{./fig/SU2_flux_b2o6_L16_hmc_Q}
\caption{$\beta=2.6$ and~$L=16$.}
\label{fig:B1c}
\end{subfigure}
\hspace{6mm}
\begin{subfigure}{0.45\columnwidth}
\centering
\includegraphics[width=\columnwidth]{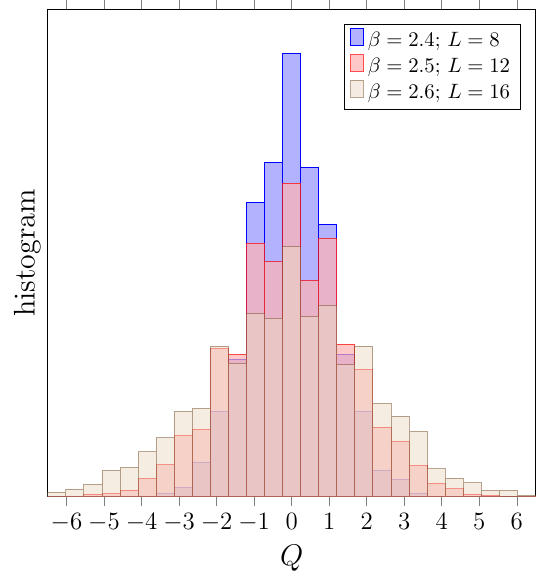}
\caption{The histogram of~$Q$.}
\label{fig:B1d}
\end{subfigure}
\caption{The HMC histories and the histogram of~$Q$, for the
$SU(2)/\mathbb{Z}_2$ theory (i.e., with the $B$-field).}
\label{fig:B1}
\end{figure}

\begin{figure}[htbp]
\centering
\begin{subfigure}{0.45\columnwidth}
\centering
\includegraphics[width=\columnwidth]{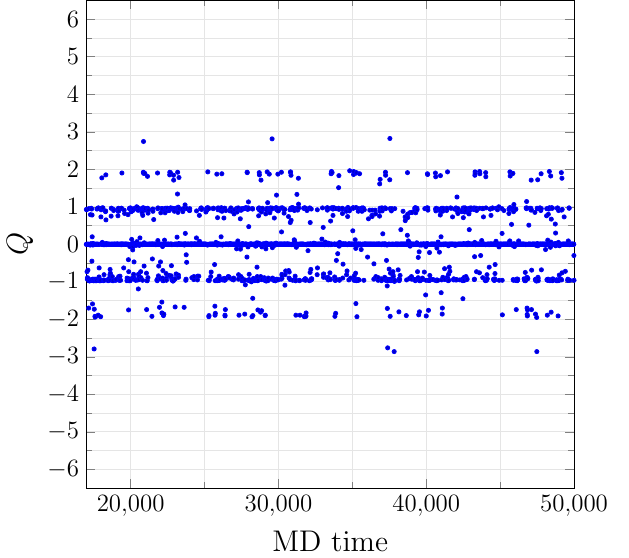}
\caption{$\beta=2.4$ and~$L=8$.}
\label{fig:B2a}
\end{subfigure}
\hspace{6mm}
\begin{subfigure}{0.45\columnwidth}
\centering
\includegraphics[width=\columnwidth]{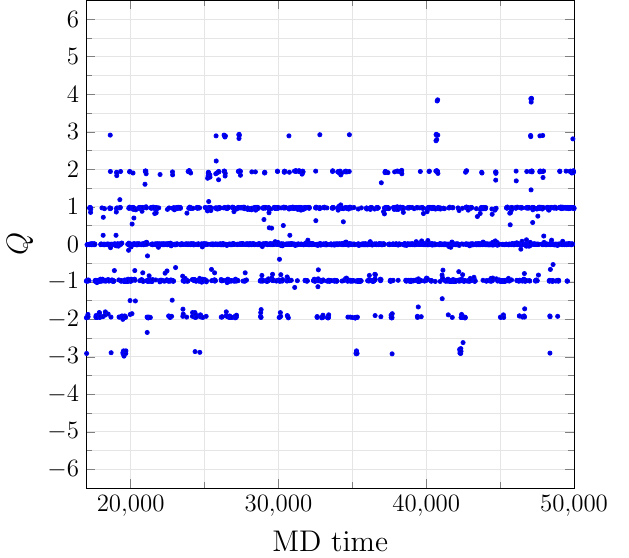}
\caption{$\beta=2.5$ and~$L=12$.}
\label{fig:B2b}
\end{subfigure}
\\[8mm]
\begin{subfigure}{0.45\columnwidth}
\centering
\includegraphics[width=\columnwidth]{./fig/SU2_noflux_b2o6_L16_hmc_Q}
\caption{$\beta=2.6$ and~$L=16$.}
\label{fig:B2c}
\end{subfigure}
\hspace{6mm}
\begin{subfigure}{0.45\columnwidth}
\centering
\includegraphics[width=\columnwidth]{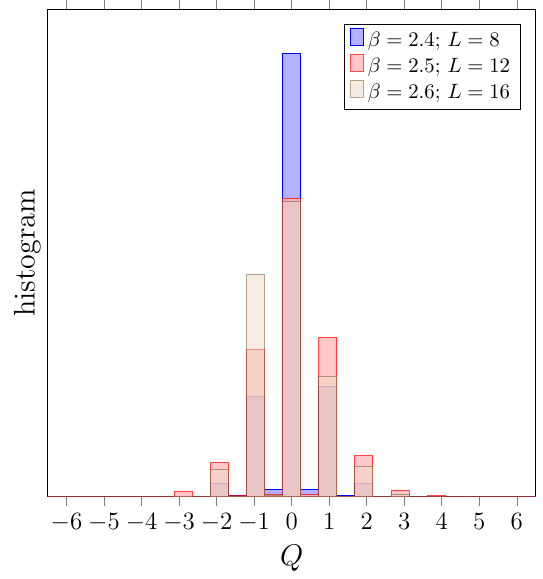}
\caption{The histogram of~$Q$.}
\label{fig:B2d}
\end{subfigure}
\caption{The HMC histories and the histogram of~$Q$, for the $SU(2)$ theory
(i.e., without the $B$-field).}
\label{fig:B2}
\end{figure}

In this appendix, we present HMC histories and histograms of the topological
charge~$Q$ in the $SU(2)/\mathbb{Z}_2$ and $SU(2)$ theories in
Figs.~\ref{fig:B1} and~\ref{fig:B2}, respectively.



%



\let\doi\relax









\end{document}